\documentclass[a4paper,english,useAMS,usenatbib]{mn2e}
\usepackage[T1]{fontenc}
\usepackage[latin9]{inputenc}
\usepackage{amstext}
\usepackage{amssymb}
\usepackage{graphicx}

\makeatletter

\title[Galaxies nurtured by mature black holes]{Galaxies nurtured by mature black holes}
\author[Masahiro morikawa]{Masahiro morikawa\thanks{E-mail:
hiro@phys.ocha.ac.jp} \\
Department of Physics, Ochanomizu University\\ 2-1-1 Otsuka, Bunkyo, Tokyo 112-8610, Japan\\
}

\makeatother

\usepackage{babel}
\begin{document}

\date{Accepted 20XX . Received 20XX}

\maketitle
\pagerange{\pageref{firstpage}--\pageref{lastpage}} \pubyear{2012}

\label{firstpage}
\begin{abstract}
Supermassive black holes (SMBH) of size $10^{6-10}M_{\odot}$ are
common in the Universe and they define the center of the galaxies.
A galaxy and the SMBH are generally thought to have co-evolved. However,
the SMBH cannot evolve so fast as commonly observed even at redshift
$z>6$. Therefore SMBH must form first before galaxy. Our goal is
to clarify how this mature SMBH forms galaxy. Furthermore we clarify
the mechanism how the SMBH designs variety of structures of galaxies.
We explore a natural hypothesis that the SMBH has been formed mature
at $z\approx10$ before stars and galaxies. The SMBH forms energetic
jets and outflows which trigger massive star formation in the ambient
gas. They eventually construct globular clusters and classical bulge
as well as the body of elliptical galaxies. We propose simple models
which implement these processes along with the standard $\Lambda$CDM-model.
We point out that the globular clusters and classical bulges have
a common origin but are in different phases. The same is true for
the elliptical and spiral galaxies. Physics behind these phase division
is the runaway star formation process with strong feedback to SMBH.
This is similar to the forest-fire model that displays self-organized
criticality. Finally we speculate several observational predictions
that may help to test the present arguments. 
\end{abstract}

\section{Introduction}

Almost all the galaxy harbors a supermassive black hole (SMBH) of
mass $M_{BH}=10^{6-10}M_{\odot}$ in its center (\cite{b15}), where
$M_{\odot}\approx2\times10^{30}$kg is the solar mass. The mass of
the SMBH is observed to have firm correlations with the basic components
of the galaxy. For example in the case of classical bulge (CB), its
mass $M_{CB}$ has the relation $M_{BH}\approx10^{-3}M_{CB}$ (\cite{b9}).\footnote{We do not consider the similar structure called pseudo-bulge which
is more like disks of spiral galaxies(\cite{b11}). }. In the case of globular clusters (GC), the number of them $N_{GC}$
in a galaxy has the relation $M_{BH}\approx10^{5.5}N_{GC}M_{\odot}$
(\cite{b11}). Therefore it might be natural to think that the galaxy
and its SMBH have co-evolved in their lives.

However, the standard coagulation of the stellar size black holes
to form SMBH within a limited time scale turns out to be very difficult
(\cite{b22}). This difficulty has become prominent by the recent
observations (\cite{b19,b25}) that report many mature SMBH exist
at around $z\approx6$ (only $7.4\times10^{8}$years after Big Bang).
Various attempts to form these SMBH through the co-evolution with
galaxies seem to be unnatural (\cite{b17}), without assuming seed
black holes of huge mass $10^{6}M_{\odot}$. Moreover there seems
to be an evidence that the above black hole/bulge mass relation $M_{BH}\approx10^{-3}M_{CB}$
has already been established or even the coefficient increases toward
the past (\cite{b24}). All of these facts strongly suggest that the
SMBH are primordial.

Therefore in this paper, we start from the hypothesis that mature
SMBH have been formed at around $z\approx10$ on top of the standard
$\Lambda$CDM model. Then the formation of all the structures, \textit{i.e.}
the stars, globular clusters, and bulges, should be attributed to
this SMBH. These structures will quickly feed back to the SMBH. Thus
the galaxy-SMBH co-evolution would have taken place in the very early
stage and the several correlations above (\cite{b9,b11}) would have
already been established then.

The early formation of SMBH at around $z\approx10$ has been studied
in the model that the Bose-Einstein condensation of the self-interacting
boson fields forms the dark energy (\cite{b21,b6,b7,b8}). In this
model, the unstable uniform condensation of the field can collapse
to form SMBH everywhere in the Universe. This is possible because
the coherent condensation does not have velocity dispersion which
prevents the collapse. Furthermore in this model, the non-condensate
component of the boson gas can contribute as the thermal dark matter
around the SMBH.

The gravitational potential of this dark matter attracts baryon gas,
and the SMBH will form a strong jet beyond the present size of the
galaxy. This is plausible from the many observations at present although
the jet ejection mechanism is not at all clear so far (\cite{b4}).
This jet will compress the surrounding gas and trigger massive star
formation there especially in the dense gas environment (\cite{b23,b18,b26,b27}).
These first stars will build globular clusters (GC), the classical
bulges (CB) in spiral galaxies (SG), and the main body of the elliptical
galaxies (EG). This physics is closely related with the mechanism
in (\cite{b14}) which displays the correlation between the SMBH mass
$M_{BH}$ and the velocity dispersion $\sigma$ of the stars in CB.
This scenario is briefly explained in section 2.

According to the above scenario, the stars of GC and CB in the spiral
galaxies, would have the same origin and are indistinguishable with
each other. On the other hand, their clustering features are different;
GC are extended in halo and CB forms the core of the galaxy. This
divide may come from the two distinguished flows of gas in the primordial
galaxy. We clarify these flows and the origin of distinct clustering
features in section 3.

Furthermore an apparent similarity of CB (in SG) and EG suggests that
the EG and SG have the common origin. The fact that SG is smaller
than EG, in average, suggests that the strength of the jet from SMBH
or the mass of SMBH are thought to be the major factor to distinguish
SG and EG. We will find much interesting parameter which clearly distinguish
SG and EG in our model described in section 4.

We try to find the simplest model extracting the most relevant physics
from very complicated galaxy formation processes. Limitations and
prospects of our approach are described in the final section 5.

\section{Early formation of supermassive black holes}

We briefly examine a possible origin of the primordial SMBH which
was formed first before any other components of the galaxy. It is
clear that the coagulation of the stellar size black holes to form
SMBH takes too long time, more than the dynamical relaxation time
scale $\tau_{rel}=\sigma^{3}/(G^{2}m\rho lnN)$, which turns out to
be $2\times10^{14}$years, where $\sigma,\:m,\:\rho,\:N,\:G$ are
the velocity dispersion, mass, mass density, number of the black holes,
and the gravitational constant, respectively. To bring this time scale
within the cosmic age $1.38\times10^{10}$years, we need seed black
holes of mass $10^{5}M_{\odot}$ to start with. Furthermore dark matter
gas is hopeless to collapse into SMBH because of its velocity dispersion
and the angular momentum. The only possibility will be the collapse
of the condensed field whose uniform component forms the dark energy
(\cite{b21,b6}). This is possible if the dark energy is the Bose-Einstein
condensation of fields (mass $m$) with an attractive self-interaction($\lambda<0$
). The condensate is characterized by the classical scalar field $\Psi(t,r)$
in the metric, assuming spherical symmetry, 
\begin{equation}
ds^{2}=\alpha^{2}\text{dt}^{2}-a^{2}\text{dr}^{2}-r^{2}\text{d\ensuremath{\theta}}^{2}-r^{2}\text{sin\ensuremath{\theta}}^{2}\text{d\ensuremath{\phi}}^{2}
\end{equation}
and obeys the equation of motion, 
\begin{eqnarray}
ra\Psi'\dot{\Psi}-2\dot{a}=0,\\
-2a'\alpha^{2}-2a\alpha\alpha'+ra\alpha^{2}\Psi'^{2}+ra^{3}\dot{\Psi^{2}}=0,\nonumber \\
\frac{2a\left(r\alpha'+\alpha\right)-2ra'\alpha}{r^{2}a^{3}\alpha}-2m^{2}\Psi^{2}+\lambda\Psi^{4}-\frac{2}{r^{2}}=0,\nonumber \\
r\left(\begin{array}{cc}
 & \dot{a}a^{2}\alpha\dot{\Psi}+a'\alpha^{3}\Psi'\\
+ & a^{3}\left(2\alpha^{3}\Psi\left(m^{2}-\lambda\Psi^{2}\right)-\dot{\alpha}\dot{\Psi}+\alpha\ddot{\Psi}\right)
\end{array}\right)\nonumber \\
=a\alpha^{2}\left(\left(r\alpha'+2\alpha\right)\Psi'+r\alpha\Psi''\right).\nonumber 
\end{eqnarray}
This set of equations easily forms black hole even if mass-less non-interacting
case, which was often used to analyze critical behavior in the black
hole formation process (\cite{b3,b10}). It was concluded that the
resultant black hole mass shows scaling properties. On the other hand,
more realistic bound for the black hole formation comes from the quantum
fluctuations. However, it turns out that even the quantum fluctuations
cannot prevent the collapse of the condensation to black hole if the
mass of the boson field exceeds the Kaup limiting mass (\cite{b13})
\begin{equation}
M_{kaup}=0.633\frac{\hbar c}{Gm}\approx\frac{m_{pl}^{2}}{m},
\end{equation}
where $m_{pl}$ is the Planck mass. If the boson mass is the order
of the present dark energy $0.01eV$, then the limiting mass becomes$1.7\times10^{22}$kg,
almost the planet Pluto mass.

After the adiabatic collapse of the dark energy, some portion of the
condensation becomes a black hole (\cite{b3}) and some other portion
will melt to form thermal boson gas around the black hole (\cite{b21}).
This latter melting process of the condensation depends on many complex
conditions and has not yet been clarified. However it would be natural
to suppose that it settles down to the thermal equilibrium of mass
density $\rho(r)=\rho_{0}(r_{0}/r)^{2}$ where $r$ is the distance
from the central SMBH and $\rho_{0},\:r_{0}$ are constants. Then
this thermal uncondensed gas behaves as dark matter (\cite{b21,b6})
since it yields the commonly observed flat rotation curve.

Suppose that the SMBH thus formed at $z\approx10$ is already surrounded
by the thermal gas of dark matter with well developed gravitational
potential. Then baryons are attracted within the free fall time scale
about $5\times10^{7}$ years. At the same time the SMBH would yield
energetic jets, which may trigger massive star formation along it
through the ram pressure of the bow shock (\cite{b23,b18}). However
the effect of the jet for star formation is not yet fully understood
as well as the jet formation mechanism itself. We do not go deep into
these problems in this paper. If the jet conveys the momentum to the
ambient baryon gas, then the pressure compresses the gas to promote
the star formation. On the other hand, if the jet conveys the energy
to the gas, then the heat makes the gas expand to prevent the star
formation. Among numerous arguments on the both directions, there
seems to be a plausible direction (\cite{b26,b27}) that promotion
and prevention of star formation coexist depending on the parameters
such as the gas density and inhomogeneity. In general the dense environment,
such as in the early stage of the Universe or in the center of the
gravitational potential, prefers the promotion of star formation.
On the other hand the dilute environment prefers the prevention and
disperses the gas clouds. 

All the above arguments are for a steady jet with fixed direction.
If the jet changes its direction rapidly less than the free fall time
scale, as we will argue in the next section, then the trajectory envelope
of the jet will form a superposed shock wave shells. A similar argument
appears in (\cite{b14}) which displays the formation of the correlation
between the mass $M_{BH}$ of the SMBH and the velocity dispersion
$\sigma$ in galactic bulge. If the typical outflow radiation balances
in momentum with the ambient gas of size $R$, then we have $GM_{b}M_{tot}/R^{2}=L_{Edd}/c$,
where $M_{b}=fM_{tot}$ is the baryon mass of the galaxy,\textit{
i.e.} the fraction $f$ of the total mass $M_{tot}$. $L_{Edd}=4\pi GM_{BH}c/\kappa$
is the Eddington limiting luminosity where $\kappa$ is the electron
scattering opacity. Then using the virial equilibrium relation $\sigma^{2}=GM_{tot}/R$,
we have (\cite{b14}), 
\begin{equation}
M_{BH}=\frac{f\kappa\sigma^{4}}{4\pi G^{2}}\propto\sigma^{4}\label{eq:mom}
\end{equation}
This successfully describes the observations (\cite{b9}). On the
other hand if we supposed energy balance, we have an extra factor
$(\sigma/c)$ on the right hand side of Eq.(\ref{eq:mom}), and $M_{BH}\propto\sigma^{5}$.
This predicts too small mass of SMBH and conflicts with observations.
Thus the observed $M_{BH}-\sigma$ relation is consistent with the
momentum balance, which suggests the jet/outflow-induced star formation.
As argued in the above, this star formation must have taken place
in the early stage of the Universe. This burst-mode star formation
induced by energetic jets should be distinguished from the spontaneous
mild star formation later time when the jets generally expel and heat
up the ambient gas to prevent the burst-model star formation.

\section{Separation into globular clusters and ellipsoids}

Now we examine how SMBH triggers star formation and makes basic components
of galaxy. The first stars, as well as their direct descendants formed
after the first supernova explosions, are thought to be the main ingredients
of the galaxy components: classical bulge (CB), globular clusters
(GC) and the elliptical galaxies (EG). These components must be formed
at the same time and same mechanism since observations indicate that
all of them are composed from very old population-II stars. Then how
these separations into components are processed?

\begin{figure}
\includegraphics[width=8cm]{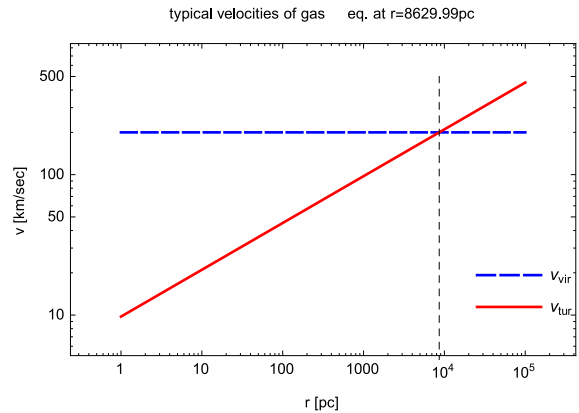}

\protect\caption{Velocity dispersion of ambient gas as a function of the distance $r$
from the central SMBH. There are two distinct velocity fields: The
constant virial velocity field $v_{in}=200km/sec$ (broken blue, dominates
inward) and the increasing turbulent velocity field $v_{out}\equiv(\epsilon r)^{1/3}$(solid
red, dominates outward). They are equal with each other at $r_{*}\approx8.6kpc$.
A star formed at $r<r_{*}$falls down toward the SMBH and forms a
bulge, while the star at $r>r_{*}$ stays far from the SMBH and forms
globular clusters. \label{fig:1}}
\end{figure}

We first consider the separation into GC and ellipsoids (\textit{i.e.
}the CB and the main body of EG). We consider that there had been
two different kinds of gas velocity fields. One is caused by the local
gravitational potential, and the other caused by the global cosmic
turbulent flow (\cite{b20})\footnote{This turbulence might be caused by the collapse of the condensation
when SMBH are formed everywhere in the Universe. }. The former velocity is given by the virial equilibrium and becomes
a constant $v_{in}\equiv(4\pi Gr_{0}^{2}\rho_{0})^{1/2}$. This is
implied by the previous isothermal distribution of dark matter density
$\rho=\rho_{0}(r_{0}/r)^{2}$. The latter is the scaling velocity
$v_{out}\equiv(\epsilon r)^{1/3},$ where $\epsilon$ is a constant.
This is derived by the scaling property of the cosmic turbulence (\cite{b20}).
This cosmic turbulence is naturally expected from the equation for
the self-gravitating fluid, whose Fourier transform

\begin{equation}
\ensuremath{\frac{{dv_{\vec{k}}^{\alpha}}}{{dt}}=-ik_{\beta}\sum\limits _{\vec{p}+\vec{q}=\vec{k}}\left({\begin{array}{cc}
\ensuremath{\left({\delta_{\alpha\gamma}-\frac{{k_{\alpha}k_{\gamma}}}{{k^{2}}}}\right)}v_{\vec{p}}^{\beta}v_{\vec{q}}^{\gamma}-\nu k^{2}v_{\vec{k}}^{\alpha}\\
+i\frac{{k^{\alpha}}}{{k^{2}}}4\pi G\delta_{\vec{k}}
\end{array}}\right)}
\end{equation}
is similar to the Smoluchowski coagulation equation. The scaling relation
derived from this equation is applied to various observations such
as the scale dependent mass density, $L/M$ ratio, magnetic field
distributions, etc. They all consistently point the value $\epsilon\approx0.3cm^{2}/sec^{3}$
(\cite{b20}). The former virial velocity dominates for $r<r_{*}$
and the latter turbulent velocity dominates for $r>r_{*}$, where
$r_{*}\approx8.6kpc$ if we assume $v_{in}=200km/sec$, as shown in
Fig. \ref{fig:1}.\footnote{This does not exclude the flat rotation curve beyond $r_{*}$. The
random nature of the turbulent flow yields fluctuations on top of
the flat rotation velocity profile produced by the relaxed dark matter.
Therefore the rotation velocity may increase or decrease outward.
However the average on these fluctuations may reproduce the flat profile
of the rotation curve. }

The baryon gas is exerted the ram pressure from the jet and forms
stars. The stars in the near region $r<r_{*}$ falls down toward the
center of the potential loosing the pressure balance and suffered
friction. Eventually they will form a big cluster that we call classical
bulge (CB). Supernova explosions and subsequent star formation may
continue. On the other hand the stars formed from the gas in the far
region $r>r_{*}$ have higher speed beyond the virial equilibrium
and have less probability to fall. Therefore these stars remain far
from the center possibly forming small clusters that we call globular
clusters (GC). 

There are some specific features of the above scenario which may be
significant in comparison with observations: (a) All the stars in
GC and CB have the same age and chemical components because they are
formed by the same gas triggered by the common jet at the same time.
(b) Each GC has almost no local angular momentum. This is because
each GC is formed in the finite region where the relative gas speed
is small and coherent according to the relation $v_{out}\equiv(\epsilon r)^{1/3}$
(Fig.\ref{fig:1}). Reflecting this small velocity dispersion at small
scale, each GC becomes compact with the typical size $r_{GC}\approx(GM)^{3/5}\epsilon^{-2/5}\approx2.5pc$
for $M=10^{5}M_{\odot}$. (c) GC are loosely bounded to the galaxy
since they tend to have higher velocities $v_{out}$ than the virial
equilibrium $v_{in}$. Therefore there may be significant number of
stray GC in between galaxies. This point should be considered in wider
viewpoint including the dwarf spheroidal galaxies\cite{b2}. In summary,
CB and GC are the same species but separated into two phases by two
distinct velocity fields.

\section{Separation into spiral and elliptical galaxies}

Next we consider the separation into the spiral (SG) and elliptical
(EG) galaxies in our scenario of mature SMBH. Apart from the disk,
the classical bulge (CB) in SG, and EG are composed from the old population-II
stars as globular clusters (GC). Therefore it would be natural to
think that CB and EG are the same species but in different phases.

In order to demonstrate this process of separation, we introduce a
simple model for the massive star formation by jets. We concentrate
on the very basic Physics behind first in order to explore the plausible
mechanism for galaxy formation at present. Therefore our model is
simply a representative one among many possible models. Suppose the
energetic jet from SMBH hits the ambient gas that is isotropically
distributed around the SMBH. We assumed that the jet triggers the
gas to form stars. Then those stars formed in the near region $r<r_{*}$
fall down toward he center. Some of them will give torque on the SMBH
through the deformation of the accretion disk. Then the jet from SMBH
changes its direction since the jet direction is thought to be parallel
to the SMBH rotation axes. In the new direction of the jet, there
will be plenty of fresh gas ready to form stars. Thus the jet will
trigger new star formation in this rich gas environment. This yields
further torque on the SMBH. This feedback dynamics will be simply
represented by the following model:

\begin{equation}
\begin{array}{cc}
\dot{g_{i}}(t)= & -\mu\left|\overrightarrow{[J}(t)]_{n}\cdot\overrightarrow{s_{i}}\right|^{\alpha}|\overrightarrow{J}(t)|g_{i}(t),\\
\dot{\vec{J}}(t)= & -\lambda[\overrightarrow{J}(t)\times\sum\nolimits _{i}\overrightarrow{s_{i}}\dot{g}_{i}(t)]_{n}\sum\nolimits _{i}\dot{g}_{i}(t)-\kappa\overrightarrow{J},
\end{array}\label{eq:4}
\end{equation}
where the vector $\vec{J}(t)$ represents the jet (direction=axes
of the jet, amplitude=strength of the jet). The scalar $g_{i}(t)$
represents the amount of gas in the $i$-th direction $\overrightarrow{s_{i}}$,
where $\overrightarrow{\{s_{i}}\}_{1\leqq i\leqq N}$ covers the whole
solid angle. The parameter $\alpha$ simply controls the beam widths
and the symbol $[*]_{n}$ represents the unit vector with the same
direction $*$. The directions $\overrightarrow{\{s_{i}}\}_{1\leqq i\leqq N}$
are designed based on the Fibonacci-Himawari coordinate system using
the golden angle $\beta=137.5^{\circ}...$ so that it covers the whole
solid angle with uniform density: $\overrightarrow{s_{i}}=\left(s\sin\gamma,\;s\cos\gamma,\;c\right)$
with $c=1-2\left(i/i_{total}\right),\:s=\sqrt{1-c^{2}},\:\gamma=\beta i$.
The first line of Eq.(\ref{eq:4}) represents the gas reduction process
by the star formation triggered by the jet. This is proportional to
the star formation rate $\mu$, the strength of the jet toward the
direction $i$ with the jet collimation parameter $\alpha$, and the
amount of gas at that direction $g_{i}(t)$. The second line represents
the time change rate of the jet. This is proportional to the feeding
efficiency $\lambda$ of the formed star to the SMBH, and the torque
exerted by the falling stars just formed. A natural fade-out term
parametrized by $\kappa$ is added.

\begin{figure}
\includegraphics[width=8cm]{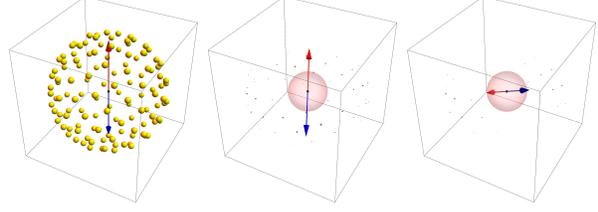}\protect\caption{The galaxy formation history in case (a), high accretion rate $\lambda=1$.
These are the snapshots of the gas distribution (small balls), jet
(arrows), and the bulge (central pink ball) derived by the numerical
calculations of Eq.\ref{eq:4}. The time flows from left to right
($t=0,\:150,\:300$). The gas was isotropically distributed at $t=0$.
Accreted objects keep exerting torque on SMBH and the jet wildly changes
its direction rapidly. This rampaging jet triggers star formation
in the whole solid angle. Eventually all the gas is exhausted and
an elliptical galaxy is left. The parameters are $\mu=1,\:n=10,\:\kappa=0.01,\:N=256$. }

\label{fig.2} 
\end{figure}

\begin{figure}
\includegraphics[width=8cm]{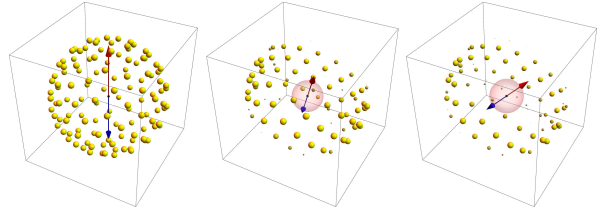}\protect\caption{The galaxy formation history in case (b), low accretion rate $\lambda=0.1$.
The same as Fig.\ref{fig.2} but with low accretion rate. The jet
is inactive and its direction does not change much. Therefore the
gas is left in a torus form almost perpendicular to the final jet
direction. The gas will eventually settle to form a disk and a spiral
galaxy is left. }
\label{fig.3} 
\end{figure}

\begin{figure}
\includegraphics[width=8cm]{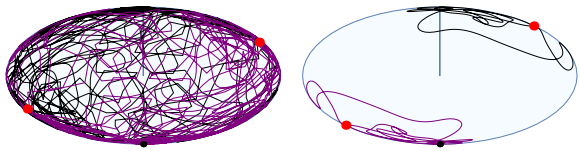}

\protect\caption{Trajectories of the moving jet-pair direction in the Mollweide projection
of the whole solid angles; horizontal and vertical axes respectively
represent the longitude and latitude. Small black and big red point
pairs represent the initial and final jet directions, respectively.
(left) The jet trajectory for the case with large accretion rate ($\lambda=1$).
The jet runs violently in the whole solid angle. This is the case
of elliptical galaxy. (right) The same but small accretion rate ($\lambda=0.055$).
The jet runs gentle and eventually the direction is settled. This
is the case of spiral galaxy. \label{fig:4}}
\end{figure}

There are two typical cases in this model(Figs.\ref{fig.2}, \ref{fig.3}).
(a) The jet is active and wildly changing its direction until finally
all the ambient gas is exhausted to form stars (Fig. \ref{fig.2}).
(b) The jet is less active and the direction is not wildly changed.
There are finite remaining gas that failed to form stars and distributed
almost perpendicular direction to the jet (Fig. \ref{fig.3})\footnote{If the settled jet-pair happens to point toward the gas remaining
regions, then the small scale star formation activity may be still
triggered which further induce density wave arm structure emanating
from there. }. In the former case (a), the resultant structure is a big star cluster
as well as small clusters GC around it, both composed from the stars
of the same age. No gas is left. This is the typical elliptical galaxy.
On the other hand in the latter case (b), the structure is a central
star cluster CB as well as GC around it, both in the gas still remaining.
Some portion of this gas located inward would eventually relaxed to
form regular disk structure, in which new stars are going to be formed
spontaneously. The remaining gas outward would be scattered and lost
by the turbulence. This is the typical spiral galaxy. Thus the elliptical
and spiral galaxies born at the same time by the same mechanism. Only
the degree of jet activity separates them into the two phases SG and
EG. The jet direction trajectories in both cases are compared in Fig.\ref{fig:4}.

\begin{figure}
\includegraphics[width=8cm]{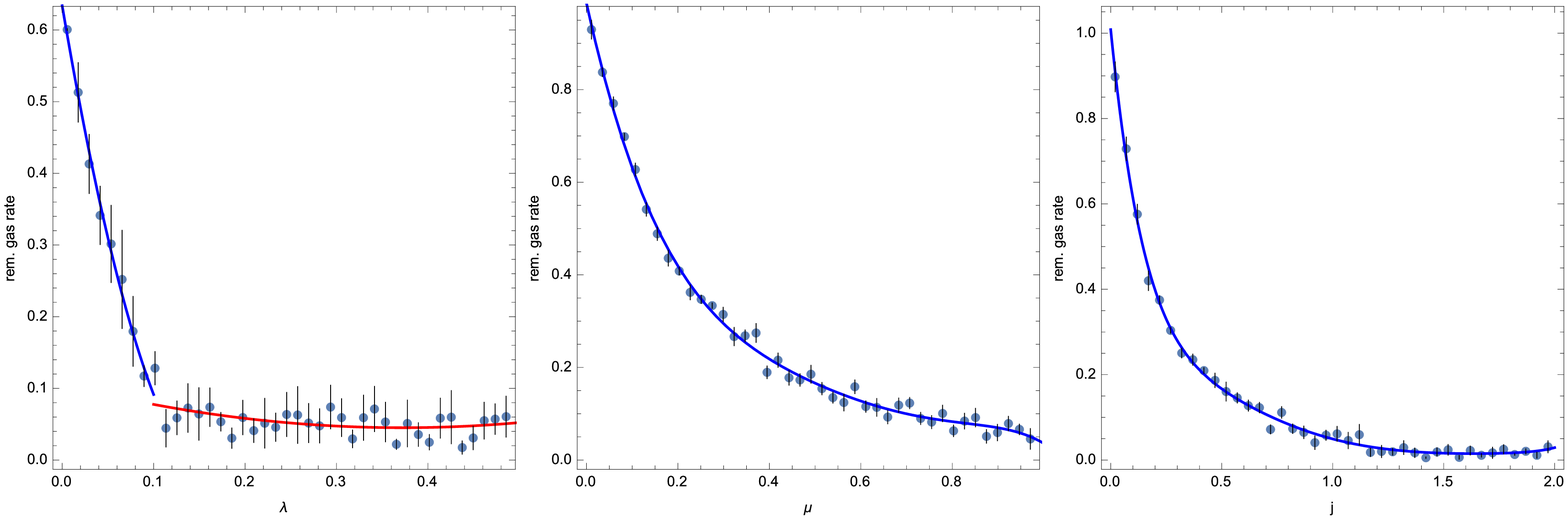}

\protect\caption{The fraction of the remaining gas that didn't form stars at $t=300$.
Each parameter is changed from the standard set $\lambda=1,\:\mu=1,\:j=1$.
(left) The case the accretion rate $\lambda$ is changed. The remaining
gas fraction has a sharp bend only in this case. (center) The case
the star formation rate $\mu$ is changed. (right) The initial jet
strength $j$ is changed. If the jet activity does not exist at all
($j=0$), it may yield a bulge-less spiral galaxy. The other parameters
are the same as in Fig. \ref{fig.2}.}
\label{fig:5} 
\end{figure}

The distinction of the two phases seems to be sharp if parametrized
by the accretion rate $\lambda$, but not by the other parameters,
as shown in Fig.\ref{fig:5}. If $\lambda$ is large, then the torque
exerted from the falling stars strongly changes the jet direction
so that more star formation takes place in the fresh ambient gas.
This further exert strong torque and this runaway continues until
all the gas is exhausted. On the other hand if $\lambda$ is small,
then the falling stars exert only weak torque to change the jet direction.
Then the jet cannot hit the sufficient amount of fresh gas to yield
torque and eventually the jet direction is settled, leaving the torus
shape gas distribution around the central bulge. The positive feedback
eventually stops, and no runaway takes place.

This galaxy formation process has similar physics to the forest-fire
model in complex systems (\cite{b12,b1}). This model is composed
of many trees on a two-dimensional lattice. Each tree on a lattice
site has a finite probability $p$ to ignite spontaneously. If ignite,
the fire spreads to the neighboring trees with some probability. On
the site of the burnt tree, a new tree has a chance to grow with some
probability. The main feature is that the fire-fire correlation length
$\xi(p)$, depends on $p$, determines the asymptotic two distinct
states: (a) The fire dies out if $\xi(p)$ exceed the system size
$L$. (b) The fire is sustained if $\xi(p)<L$. Percolation caused
by the positive feed back or runaway process is the common feature
in our galaxy model and the forest-fire model although the former
is deterministic, as Eq.(\ref{eq:4}), and the latter is probabilistic.

In our model, the parameter of accretion rate $\lambda$ may particularly
be important to divide the galaxies into the elliptical and spiral
(Fig.\ref{fig:5} left). The special value $\lambda_{*}\approx0.1$
divides the EG ($\lambda>\lambda_{*}$) and SG($\lambda<\lambda_{*}$).
This parameter corresponds to the relevant parameter $p$ in the above
forest-fire model. 

\begin{figure}
\includegraphics[width=8cm]{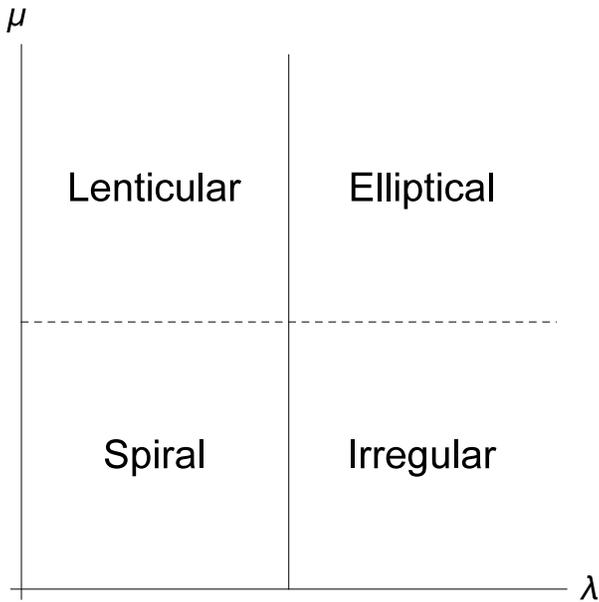}

\caption{Possible separation of galaxy species according to the parameters
$\lambda,\mu$ --- a schematic diagram. Elliptical galaxies are from
large $\mu$, $\lambda$, spirals are from small $\mu$, $\lambda$,
lenticular galaxies are from large $\mu$ and small $\lambda$, and
irregular galaxies from small $\mu$ and large $\lambda$. Note that
these classifications are for the galaxies just formed, and does not
specify the classifications at present $z=0$. Many irregular galaxies
might have changes their form into spirals, for example (see the test).
\label{fig:6}}
\end{figure}

It may be interesting to examine a possible galaxy classification
further in our model, focusing on the representative parameters $\mu$
and $\lambda$ (Fig.\ref{fig:6}). Large star formation rate $\mu$
is provided, for example, by the dense ambient gas such as in the
bottom of the gravitational potential produced by the huge dark matter
halo. On the other hand large feeding efficiency $\lambda$ is provided,
for example, by small angular momentum of the whole gas cluster. Therefore
the galaxies produced in dense ambient gas with small angular momentum
correspond to large $\mu$ and large $\lambda$. This set of parameter
provides strong and violent jet and leaves a big cluster of stars
without remaining gas. This may yield elliptical galaxies. Contrary
the galaxies produced in dilute ambient gas with large angular momentum
correspond to small $\mu$ and small $\lambda$. This set of parameter
provides weak and steady jet and leaves small cluster of stars with
plenty of gas remaining. This may yield spiral galaxies. The intermediate
case that the galaxies produced in dense ambient gas with large angular
momentum correspond to large $\mu$ and small $\lambda$. This set
of parameter provides strong jet and leaves a big cluster of stars
with some amount of remaining gas. This may yield lenticular galaxies.
The remaining case of small $\mu$ and large $\lambda$ would not
yield any prominent regular structures. Smallest values of $\mu$
and $\lambda$ may yield apparently tiny galaxies. These tiny galaxies
would be clearly distinguished from GC by the existence of their central
SMBH and dark matter, which are absent in GC (\cite{b28}). 

According to our scenario, these galaxy species are \textit{intrinsic
}and do not change in time. However the ratio of galaxy species may
be slightly altered by later merger processes, which may not dominate
nor are relevant in our scenario though. Moreover we can speculate
the \textit{apparent} evolution of the galaxy species. The produced
EG has no gas and therefore no further prominent star formation process,
except some AGN in which the jet is re-activated for any reason. Thus
the population of EG species will not largely change in time. On the
other hand the SG has plenty of gas remaining around the central bulge
and the spontaneous star formation actively continues. This spontaneous-mode
of star formation is contrasted with the burst-mode of them due to
the jets. The former stars mainly form the population I and the latter
the population II. In SG, the ambient gas was initially irregular
and inhomogeneous just after the formation because of the random jets.
This primordial SG will eventually form regular disk around the central
bulge by consuming their locally excess kinetic energy to trigger
new star formation through the shocks. Therefore the apparent irregular
SG will evolve toward regular SG. These are the distinctive trend
of our scenario among other theories of galaxy formation including
the merger hypothesis.

\section{conclusions and prospects}

We proposed a scenario that the primordial supermassive black hole
(SMBH) made the galaxy. Energetic jets from the SMBH give ram pressure
to trigger the massive star formation in the primordial dense gas
environment. We focused on the following two aspects of this scenario.

The first aspect is the comparison of a globular cluster (GC) and
a classical bulge (CB). According to our scenario, they are the same
species but in distinct phases. We proposed that the difference comes
from two kinds of velocity fields of the gas around the SMBH: (a)
the virial velocity field $v=const.$ associated with the dark matter
distribution that dominates inside of the galaxy, and (b) the cosmic
turbulent field, which obeys the Kolmogorov scaling $v\propto r^{1/3}$,
that dominates outside. Some portion of the star formed by the jet
in (a) falls into the center and yields CB. On the other hand the
star formed in (b) stays in halo and yields compact GC reflecting
the characteristic Kolmogorov velocity field. Thus the two species
CB and GC are formed simultaneously from the same gas, but in distinct
velocity fields.

The second aspect is the comparison of the spiral galaxies (SG) and
the elliptical galaxies (EG). According to our scenario, CB (in SG)
and EG are the same species but in distinct phases. We proposed that
the difference comes from the amount of jet activity. We introduced
a simple model that describes this discrimination. According to this,
(a) large star formation rate $\mu$ (for example in the dense gas
environment) and large feeding efficiency $\lambda$ (for example
in the case of small angular momentum) make strong feedback torque
on SMBH that causes runaway flipping of the jet direction in the whole
solid angle. Finally all the gas is exhausted and an EG is left. On
the other hand (b) small $\mu$ and $\lambda$ make weak torque and
the jet direction change mildly without runaway. This makes smaller
cluster in the center (CB) and the leftover gas will eventually form
a disk, leaving a SG. Thus the existence/absence of the runaway separates
the same species into two phases. This process has the common physics
to the forest-fire model in complex systems.

Thus a SMBH defines the center of a galaxy and various star clusters
in the galaxy (GC, CB, EG) are nurtured by the SMBH through the energetic
jet emanating from the SMBH. According to this scenario, the hypothesis
of the population-III stars that formed spontaneously may not be necessary.

Our simple analysis may be a useful supplement to the solid simulations
of galaxy formation including all physical processes. We have ignored
many detail dynamics such as the star formation process, back reaction
to the jet, jet formation and collimation, and transfer of the angular
momentum, etc. We wanted to extract the most relevant physics in the
complex galaxy formation process and to construct a natural model
based on them. We would like to report how the elaboration of our
model including the above fundamental processes can be possible in
our future study. 

Several final remarks are in order. The jet activity may leave its
trace in the faint structures in a spiral galaxy. The final stage
of the jet would be gentle and the jet will generally have small precession.
This jet may form a double-cone shape region in which the stars are
massively formed. Subsequent supernova explosions of those stars may
leave high energy electrons and protons there captured and stored
in the magnetic fields. This relic ionized region will be observed
in the early galaxies. However this relic structure may be contaminated
with the relatively recent jet activity which often observed (\cite{b5}).

In our scenario, a SMBH always defines the center of the galaxy. Therefore
even the expelled SMBH, after multiple merger of galaxies, if any,
will form a new galaxy independently around it. In any case, galaxy
merger is not a dominant process in our scenario. Rather, we expect
a cluster of SMBH which will be formed by the instability of the huge
condensed field. In this case, multiple SMBH are expected to form
on a plane after the collapse of the condensation in the form of pancake.
This process does not destroy any of the $\Lambda$CDM model in which
the dark matter forms individual gravitational potential. Actually
we simply need a tiny fraction of those dark matter cluster becomes
condensed and collapse to form SMBH. 

We hope we will soon be able to report these considerations by checking
the fundamental processes theoretically and observationally.

\section*{Acknowledgments}

The author would like to thank Hideaki Mouri, Akika Nakamichi, Tohru
Tashiro, and all the laboratory members of Ochanomizu University for
fruitful discussions and critical comments. 

\label{lastpage}
\end{document}